\documentclass[preprint,12pt]{elsarticle}

\usepackage{amssymb}
\usepackage{multicol}
\usepackage{subcaption}
\usepackage{float}
\usepackage{siunitx}
\usepackage{lipsum}
\makeatletter
\def\ps@pprintTitle{%
 \let\@oddhead\@empty
 \let\@evenhead\@empty
 \def\@oddfoot{}%
 \let\@evenfoot\@oddfoot}
\makeatother
\usepackage{graphicx}
\usepackage[export]{adjustbox}

\journal{Nuclear Physics B}

\begin{document}

\begin{frontmatter}



\title{TRAJEDI: Trajectory Dissimilarity}


\author{Pedram Gharani\fnref{label1}} 
\ead{peg25@pitt.edu}
       
\author{Kenrick Fernande\fnref{label2}} 
\ead{kenrick@cs.pitt.edu}

\author{Vineet Raghu\corref{cor1}\fnref{label2}} 
\ead{vkr8@pitt.edu}

\fntext[label1]{School of Computing and Information, Department of Informatics and Networked Systems, University of Pittsburgh}

\fntext[label2]{School of Computing and Information, Department of Computer Science, University of Pittsburgh}

\begin{abstract}
The vast increase in our ability to obtain and store trajectory data necessitates trajectory analytics techniques to extract useful information from this data. Pair-wise distance functions are a foundation building block for common operations on trajectory datasets including constrained SELECT queries, k-nearest neighbors, and similarity and diversity algorithms. The accuracy and performance of these operations depend heavily on the speed and accuracy of the underlying trajectory distance function, which is in turn affected by trajectory calibration. Current methods either require calibrated data, or perform calibration of the entire relevant dataset first, which is expensive and time consuming for large datasets. We present TRAJEDI, a calibration aware pair-wise distance calculation scheme that outperforms naive approaches while preserving accuracy. We also provide analyses of parameter tuning to trade-off between speed and accuracy. Our scheme is usable with any diversity, similarity or k-nearest neighbor algorithm.
\end{abstract}

\begin{keyword}
Trajectory\sep similarity\sep dissimilarity



\end{keyword}

\end{frontmatter}


\section{Introduction}
Ubiquitous mobile devices are able to collect various types of data especially motion-related one such as location, inertial, or even visual data. There are wide veritey of analyses and applications which are defined and conducted based on the availability of these types of data such as \cite{gharani2017artificial}, \cite{suffoletto2018using}, \cite{gharani2017context}. Spatial data has enjoyed increasing recognition for its uniqueness over the past few decades. Over the past decade in particular, the rise in location-enabled devices, smart or otherwise, has led to an increase in the spatial and trajectory data available. The corresponding rise in cheaply-available computational power and storage capacity has been exploited by spatial databases such as T-Drive \cite{yuan2010t},\cite{yuan2011driving} and GeoLife \cite{zheng2010geolife}. These are however just the raw ingredients needed to extract information from the data. Trajectory data, our focus in this paper, can provide valuable insights in a variety of scenarios ranging from business advertising and recommendation \cite{liu2015location} and geo-social media \cite{bao2012location} to disaster planning \cite{rao2012spatiotemporal} and green commuting route decision making \cite{shen2013investigating},\cite{chen2011discovering}.

A variety of trajectory storage and analysis techniques have already been proposed in the literature to, for example trajectory pattern mining \cite{giannotti2007trajectory}, unveiling the complexity of human mobility \cite{giannotti2011unveiling},trajectory search \cite{chen2010searching}, semantic query of trajectory\cite{bogorny2009st}, and similarity search \cite{chen2005robust}. Distance measures for similarity/dissimilarity are a key building block for a variety of trajectory analysis algorithms, such as diversity \cite{yin2011diversified}  and k-nearest neighbors \cite{yu2005monitoring}. Currently, Dynamic Time Warping \cite{vlachos2002discovering} and Synchronized Euclidean Distance \cite{potamias2006sampling} enjoy acceptance and popularity as distance measures for trajectories. However, a key real world challenge often not addressed in this context is the variable sampling rate for available trajectory data. Since trajectory data comes from a variety of sources, simply cleaning data is insufficient for the preprocessing phase. To address this challenge, the literature contains work on calibration methods for trajectory data\cite{su2015calibrating}.

Calibration, while useful, is an expensive operation. Our experimental evaluations show that state-of-the-art grid based calibration, proposed in \cite{su2015calibrating} has nearly exponential growth in cost for increase in the number of points in a trajectory. To put this in context, popular open-source trajectory datasets such as T-Drive and GeoLife contains hundreds of thousands of points in each trajectory. Current works assume the availability of calibrated data, or perform calibration prior to analysis\cite{su2015calibrating}. This approach is feasible for small data sets, but will not scale to larger datasets. This problem will only be aggravated by increasing dataset size and real-time analysis requirements.

To tackle this issue, we propose a calibration-aware distance measurement algorithm TRAJEDI in this paper. TRAJEDI selectively calibrates portions of trajectories in a result set to obtain pair-wise distances with better response times and comparable accuracy. We implement and evaluate TRAJEDI with synthetic data to demonstrate the feasibility of our scheme, as well as the effects of tuning parameters to control the trade-off between response time and accuracy. Our main contributions include:

\begin{itemize}
\item A calibration-aware distance measurement algorithm that can be used as the building block for analytics techniques such as diversity or k-nearest neighbors (Section 3)
\item Implementation and experimental evaluation of TRAJEDI, our proposed scheme, on synthetic datasets (Section 4)
\end{itemize}

\section{THE NEED FOR CALIBRATION-AWARE DISTANCE}

In this section we introduce the Dynamic Time Warping distance function, and the trajectory calibration methods we use. While these are published works, we did not find open source implementations available and hence implemented our own versions. We provide the details of these implementations in this section as well. Finally, we present results from our studies of the cost of calibration on trajectories, demonstrating the need for calibration-aware distance measurements between trajectories. Our results show that computing fully calibrated trajectories scales poorly even on relatively small synthetic data.

\subsection{Dynamic Time Warping}
Dynamic Time Warping (DTW) is a distance measurement metric originally used for time series data when the time domain itself is unimportant to the distance calculation and the time series being measured are different lengths. The metric finds the alignment between time series such that the distance between the series are minimized. Alignment in this context means the mapping of points in the first time series to points in the second time series that maintains the ordering of the points. For example, if we have time series x with points $x_1$, $x_2$, and $x_3$, and time series $y$ with points $y_1$ and $y_2$. The optimal alignment could be the map $\left(x_1 \rightarrow y_1, x_2, x_3 \rightarrow y_2\right)$; however, the map $\left(x_1,x_3 \rightarrow y_1, x_2 \rightarrow y_2\right)$ would not be valid due to the mixing of the time order of the points of $x$.

\begin{figure}[H]
\begin{center}
\includegraphics[scale=.5]{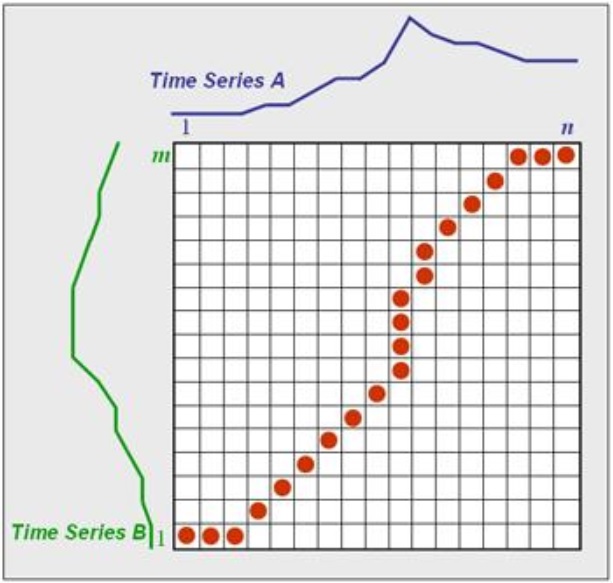}
\end{center}

\caption{Dynamic Time Warping Matrix}

\label{fig1}
\end{figure}

To efficiently compute this, we employ the standard dynamic programming algorithm which constructs a matrix $M$ of size $n \times m$, where $n$ is the number of points in the first time series, and $m$ is the number of points in the second time series. Within the matrix, an example of which is shown in Figure \ref{fig1}

\begin{equation}
M\left(i,j\right)=min\left(M\left(i-1,j-1\right),M\left(i-1,j\right),M\left(i,j-1\right)\right)+D\left(i,j\right)
\end{equation}

Where $D$ is distance function. The distance function here is defined as simple Euclidean distance between points i and j. Extending this algorithm to trajectories is done in a straightforward way, where all aspects of the algorithm remain the same, and Euclidean distance is used to compute the distance from point i to point j in the matrix computation.
\subsection{Grid-Based Calibration}
Our calibration technique is a state of the art technique that assumes no underlying properties of the dataset and is thus generally applicable. To begin, a reference set of points is constructed using a grid based approach. In this, the entire data space is divided into an $N \times N$ grid and anchor points are placed at the center of each square in the grid. These anchor points are then used as the final points that all points of the trajectories will map to after the following two phases of the algorithm: \textbf{Alignment} and \textbf{Complement}.

\begin{figure}[H]
\begin{center}
\includegraphics[scale=.75]{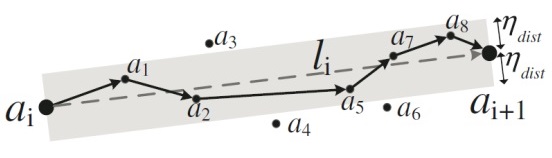}
\end{center}

\caption{Sample Grid Based Calibration}

\label{fig2}
\end{figure}

The \textbf{alignment} phase of the algorithm involves snapping trajectory points to anchor points that are close in distance. In particular, for each trajectory point, a nearest neighbor search is done on the anchor point set to find candidate anchor points that are within a system specified threshold from the current trajectory point. If no anchor points are within this threshold, then the point is removed from the set. If multiple consecutive trajectory points map to the same anchor point, then this anchor point is only treated as a single point in the final calibrated trajectory, thus reducing the dimensionality of the trajectory. A helpful visual from \cite{su2015calibrating} (Figure \ref{fig2}) explains this as well.

The \textbf{complement} phase of the algorithm involves data
interpolation between consecutive aligned anchor points following the alignment phase. For a given pair of consecutive anchor points, $a_1$ and $a_2$, the algorithm first constructs the line segment $\bar{a}$ connecting $a_1$ and $a_2$ and finds all points within a specified threshold from $\bar{a}$. Then, it iterates through the points $\left(p_1,\cdots , p_k \right)$ in the set ordered by increasing distance from $a_1$. For each point $p_i$, $p_i$ is added to the set only if the moving trend of is unchanged during this process. More concretely, the point is added only if the angle between the line $p{i−1}$ to $p_i$ and the line $\bar{a}$ is less than $\frac{\pi}{2}$.

\subsection{The Cost of Calibration}
\begin{figure}[H]
\begin{center}
\includegraphics[scale=.3]{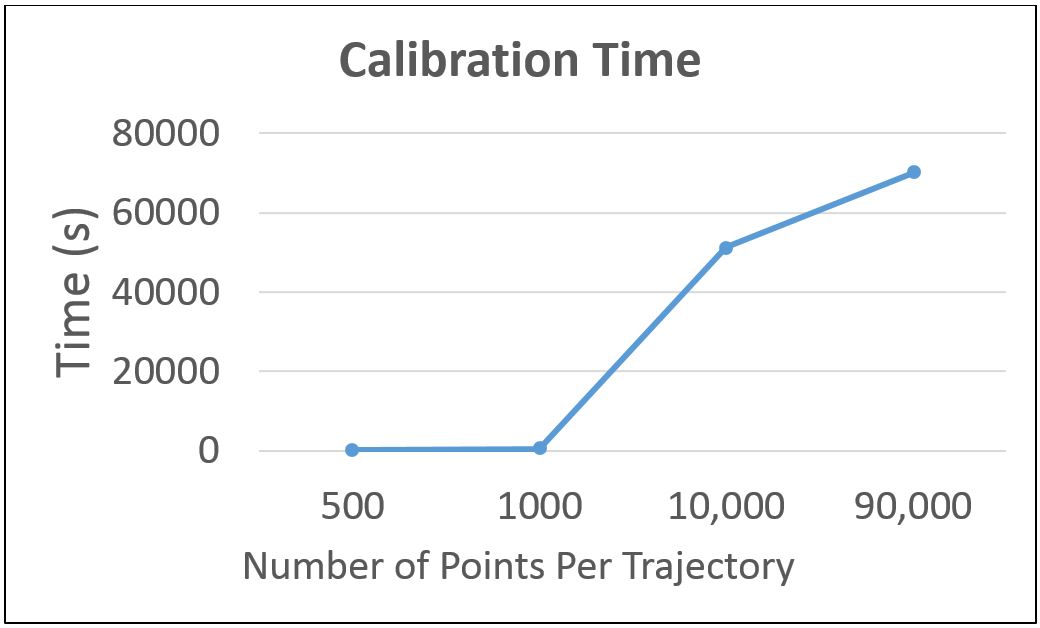}
\end{center}
\caption{Calibration Cost vs. Points per Trajectory}
\label{fig3}
\end{figure}

The issue with the state of the art calibration approach is that it is computationally expensive. Performing the nearest neighbor searches and geometric calculations result in a slow process for large datasets. To illustrate this, we examined the runtime performance of our implementation of the calibration scheme for growing subsets of the T-Drive dataset (Figure \ref{fig3}). In this figure, we see that the time to compute the fully calibrated set of trajectories becomes prohibitively expensive as the set increases. To rectify this situation, we next discuss our scheme for computing trajectory distances while keeping this cost in check.

\section{HOW TO WRITE A TRAJEDI}
In this section, we present TRAJEDI, our scheme for computing pair-wise distances between trajectories that balances accuracy and efficiency. we describe how our scheme works as a calibration aware distance function, including the system design and control flow.

\subsection{TRAJEDI scheme}
Our scheme for balancing the accuracy-efficiency trade off between fully calibrated trajectories and uncalibrated trajectories involves using a sliding window over the DTW matrix of a pair of trajectories to determine the most important region of a trajectory to calibrate. Specifically, the method examines the entries located in the diagonal of the DTW matrix at the edge of the sliding window and finds the difference between these entries. In doing so, it determines the location of the trajectory that contributes most strongly to the DTW distance of the overall matrix. The window size is controlled by a parameter $\alpha$, where $0 < \alpha < 1$ represents the percentage of the overall matrix to be calibrated. High settings of $\alpha$ represent full calibration with low efficiency while low settings of $\alpha$ represent the converse.

More formally, let $M$ be the $n \times m$ DTW matrix, where $n < m$, computed on the raw trajectories $T1$ and $T2$. Then, given input parameter $\alpha$, the evaluated windows range from point$(1,1)$, $\left( \alpha \times n , \alpha \times m \right)$ to the final window of $\left( \alpha \times n , \alpha \times m \right)$ , $\left(  n , m \right)$. The step size for window the $n$ coordinate ranges
is $1$, where the step size for the $m$ coordinate is $\frac{m}{n}$.

An issue that arises in using this scheme for all pairwise trajectory distance calculations in a full trajectory dataset is that by the end of the calculations, all trajectories are nearly fully calibrated even for small settings of alpha. This is due to the fact that the DTW calculation is repeated for each trajectory. To alleviate this issue, we insert the requirement that all trajectories will only be calibrated during one DTW calculation with another trajectory. To determine this partner for all trajectories, we suggest some potential schemes that leverage the fact that the DTW matrix is cheap to calculate.

Our baseline scheme is to randomly select one partner from all trajectories to use in the DTW calculation to calibrate the trajectory. We call this baseline \textbf{Random}. Our other schemes involve computing the whole pairwise distance matrix on the raw trajectories, and then using this information to select a partner. We select partners based upon trajectories that are furthest apart, and we call this scheme \textbf{Largest}/\textbf{Furthest}, and we select partners based upon the closest trajectories, and we refer to this as \textbf{Shortest}.

\section{EXPERIMENTAL EVALUATION}
In this section we describe how we generated the synthetic data used in the evaluation, the experimental settings, and the various measurements.
\subsection{Evaluation Data and Setting}

Our dataset was simulated by taking an anchor point reference set that was constructed using the method described in Section 2, with a $1000 \times 1000$ grid. Each trajectory began at a randomly selected anchor point and proceeded in a random walk, but only in directions that continued the moving trend of the trajectory or that resulted in no change in the trajectories motion. To simulate the issue of different sampling rates, we used a Gaussian distribution to decide how many points to remove. A small amount of Gaussian noise was injected into each point to simulate the effect of uncertainty in GPS signals.

Our small dataset consisted of 50 trajectories originally with 1500 points. The amount of points in the final version came from $N(800,200)$, which produces trajectories with significantly differing sampling rates. Our large dataset consisted of 200 trajectories with the same sampling rate values.

The entire system was implemented in Java 1.8 (front end) with PostgreSQL 9.4 serving as the database backend. For these initial experiments, all time costs reported are in terms of processing time, thus ignoring any I/O costs that may be incurred by the database implementation.

\section{Analysis of Results}
To evaluate our scheme, we measured both \textit{accuracy} and \textit{efficiency}. \textit{Accuracy} is defined as the effect on results, and \textit{efficiency} captures the total percentage of trajectories calibrated. As baselines for comparison, we use algorithms that perform no calibration and full calibration. Accuracy in these experiments refers to the normalized average difference over entries in the pairwise distance matrices over all trajectories in the dataset. For these experiments, all accuracy measurements refer to the distance in comparison to the ground truth trajectory dataset before noise is introduced and sampling rates are changed.

First, we evaluate the effects of the window size parameter
on accuracy (Figure \ref{fig5}) and efficiency (Figure \ref{fig5}) for different trajectory pair choice selections . In addition, Figure \ref{fig6} shows how changing the size parameter affects the wall clock time required for the computations to complete.

\begin{figure}[H]
\begin{center}
\includegraphics[width=\linewidth]{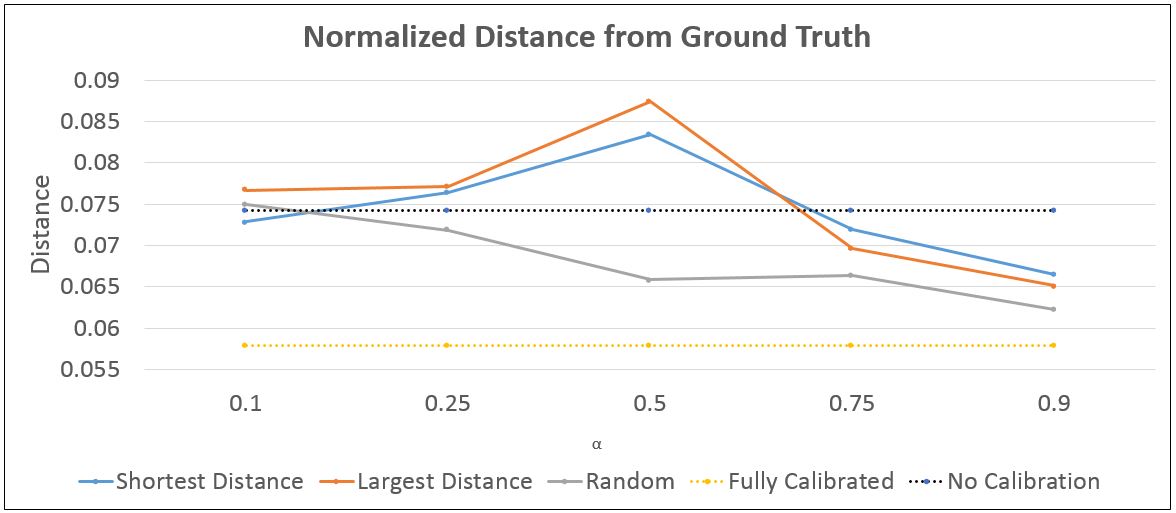}
\end{center}

\caption{Effect on \textit{Accuracy} for Window Size settings}

\label{fig4}
\end{figure}

\begin{figure}[H]
\begin{center}
\includegraphics[width=\linewidth]{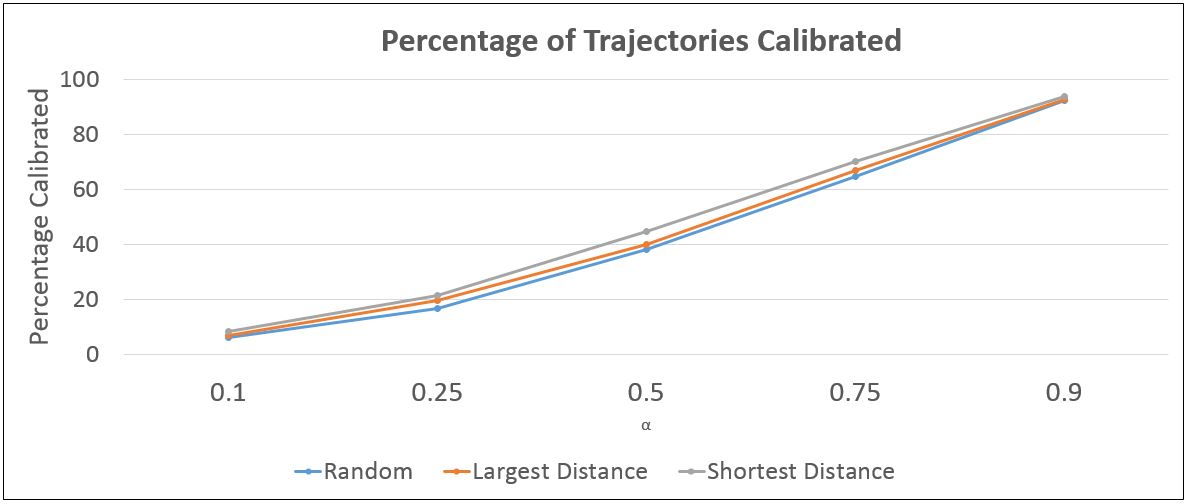}
\end{center}

\caption{Effect on \textit{Efficiency} for Window Size settings}

\label{fig5}
\end{figure}

\begin{figure}[H]
\begin{center}
\includegraphics[width=\linewidth]{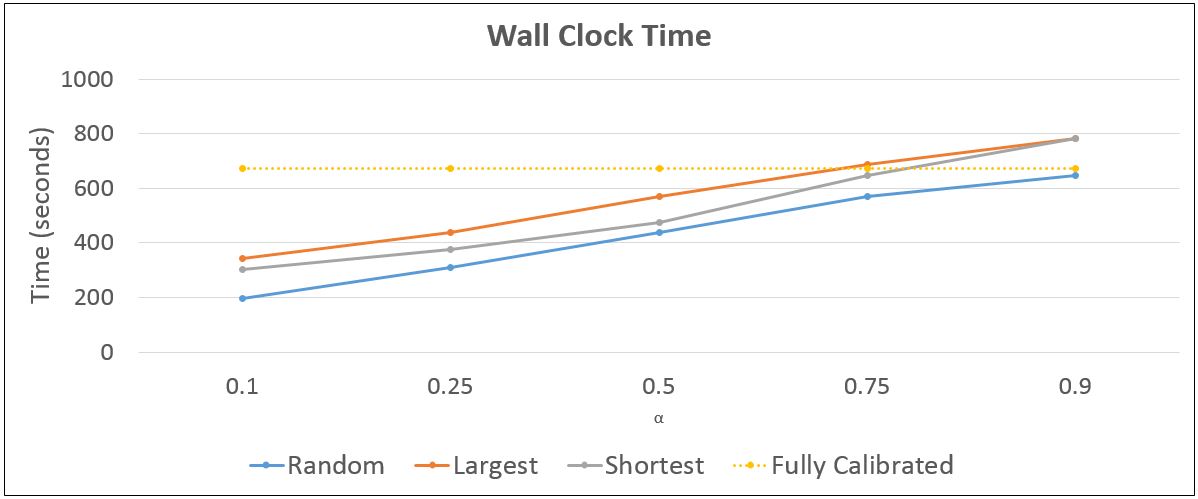}
\end{center}

\caption{Wall Clock Time Measurements}

\label{fig6}
\end{figure}

First, we evaluate the effects of the window size parameter on accuracy (Figure \ref{fig4}) and efficiency (Figure \ref{fig5}) for different trajectory pair choice selections on our small 50 trajectory dataset. The accuracy results are varied with the random choice of trajectory partner performing the best. An interesting result of these experiments is that for middling values of the calibration parameter we see a degradation of accuracy to worse than baseline levels. A possible explanation for this is that a middle window size causes poorer selection of calibration window that reduces the accuracy even though a large percent of the trajectory is calibrated. The efficiency results are as expected in that the percentage of calibrated trajectory points on average is very similar to the setting of the parameter value. In Figure \ref{fig6} we can see that in terms of absolute wall clock time our scheme delivers improvements in efficiency at the slight cost of accuracy measurements.

\begin{figure}[H]
\begin{center}
\includegraphics[scale=.375,frame]{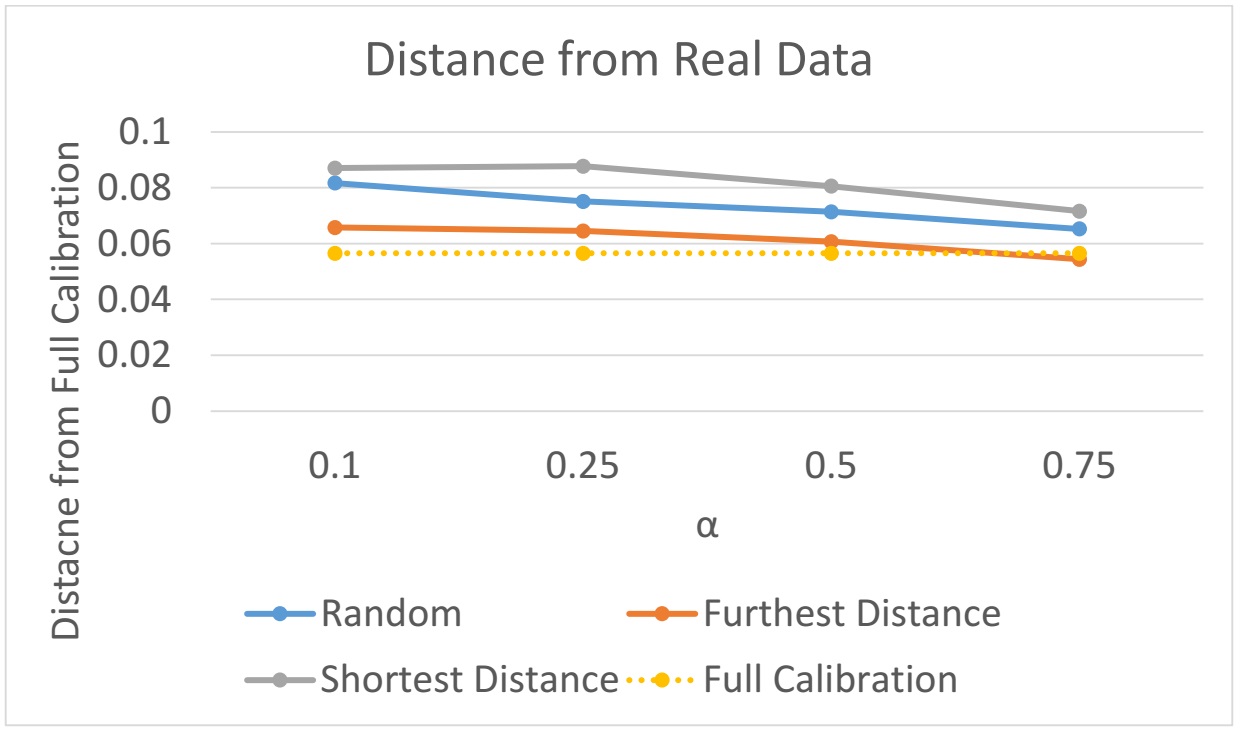}
\end{center}

\caption{Effect on Accuracy for Window Size settings}

\label{fig7}
\end{figure}

\begin{figure}[H]
\begin{center}
\includegraphics[scale=.375,frame]{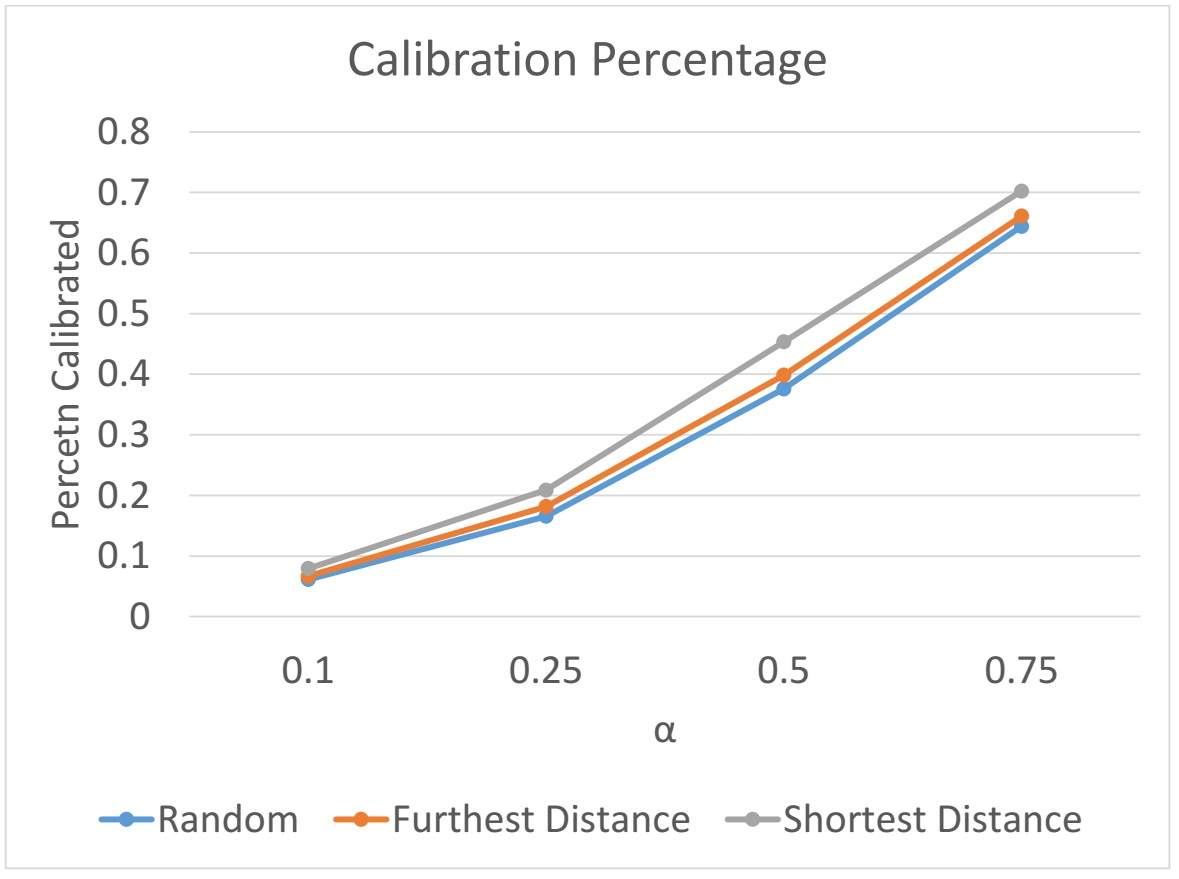}
\end{center}

\caption{Effect on Efficiency for Window Size settings}

\label{fig8}
\end{figure}

Next, we compare the performance to the larger 200 trajectory datasets in Figures \ref{fig7} and \ref{fig8}. These experiments demonstrate that the accuracy results from the smaller dataset do not necessarily generalize to other dataset distributions. In particular, we no longer see that middle values of the calibration parameter cause a degradation of performance. In addition, in these experiments we see that choosing a partner to calibrate against based on the one that is furthest away actually provides the best performance. Thus, in terms of accuracy, we conclude that the performance of these various schemes are dataset dependent and further exploration into this needs to be done to have definite conclusions. The efficiency results appear much the same as the smaller 50 trajectory dataset, though in this large dataset smaller settings of the calibration parameter result in a greater percentage of the trajectory being calibrated.

\section{CONCLUSIONS}
In this project, we proposed a calibration-aware distance measurement algorithm, TRAJEDI, which can be used for dissimilarity measurement among trajectories. TRAJEDI can effectively measure the distance between trajectories considering the most influential segment of the trajectories providing robust criterion for dissimilarity estimation of trajectories. Our evaluation demonstrates that we can trade-off between the level of desired accuracy and time performance.

Generally, measuring similarity/dissimilarity provides a useful tool for the analysis of trajectories across a wide range of applications. The results of the measurements can be used for wide range of applications such as spatial indexing, equipping analysts with more robust tool for interpretation of pedestrian, vehicle, and animal movement. Using this algorithm for measuring dissimilarity/similarity enables experts and general users of trajectory databases to explore their data more effectively in practical applications.

\bibliographystyle{elsarticle-harv}

\end{document}